\documentclass[10pt,aps,prl,twocolumn,nofootinbib,superscriptaddress,showpacs,showkeys,preprintnumbers,floatfix,byrevtex]{revtex4-1}

\usepackage{epsfig}
\usepackage{amsmath,amssymb}
\usepackage{multirow}
\usepackage{graphicx}
\usepackage[usenames,dvipsnames,svgnames,table]{xcolor} 
\usepackage[super]{nth} 
\usepackage{soul}										
\setstcolor{red}	                                                                       

\newcommand{\Ex}[2]{\ifmmode{#1\times10^{#2}}\else{$#1\times10^{#2}$}\fi}

\DeclareMathAlphabet{\mathpzc}{OT1}{pzc}{m}{it}

\newcommand{\nuc}[2]{\ifmmode{{}^{#2}\mathrm{#1}}\else{${}^{#2}\mathrm{#1}$}\fi}

\newcommand{\eV}{\,\mathrm{eV}}
\begin{document}

\title{The viability of the 3+1 neutrino model in the supernova neutrino process}

\author{Heamin Ko}
\affiliation{Department of Physics and OMEG Institute, Soongsil University, Seoul 06978, Republic of Korea}

\author{Dukjae Jang}
\email{havevirtue@ssu.ac.kr}
\affiliation{Department of Physics and OMEG Institute, Soongsil University, Seoul 06978, Republic of Korea}

\author{Motohiko Kusakabe}
\affiliation{School of Physics and International Research Center for Big-Bang Cosmology and Element Genesis,\\ Beihang University, Beijing 100083, China}

\author{Myung-Ki Cheoun}
\affiliation{Department of Physics and OMEG Institute, Soongsil University, Seoul 06978, Republic of Korea}

\date{\today}

\begin{abstract}
Adopting the 3+1 neutrino mixing parameters by the IceCube and shortbase line experiments, we investigate the sterile-active neutrino oscillation effects on the supernova neutrino process. For the sterile neutrino ($\nu_s$), we study two different luminosity models. First, we presume that the $\nu_s$ does not interact with other particles through the standard interactions apart from the oscillation with the active neutrinos. Second, we consider that $\nu_s$ can be directly produced by $\nu_e$ scattering with matter. In both cases, we find that the pattern of neutrino oscillations can be changed drastically by the $\nu_s$ in supernova environments. Especially multiple resonances occur, and consequently affect thermal neutrino-induced reaction rates. As a result, $^7$Li, $^7$Be, $^{11}$B, $^{11}$C, $^{92}$Nb, $^{98}$Tc and $^{138}$La yields in the $\nu$-process are changed. Among those nuclei, $^7$Li and $^{11}$B yields can be constrained by the analysis of observed SiC X grains. Based on the meteoritic data, we conclude that the second model can be allowed while first model is excluded. The viability of the second model depends on the sterile neutrino temperature and the neutrino mass hierarchy.   
\end{abstract}

\pacs{97.60.Bw, 14.60.St}

\maketitle

Neutrino physics has become a new sight of the astrophysics. Beginning with observations of neutrinos from supernova (SN) 1987A \cite{Hirata:1987hu}, various neutrino properties have aroused lots of intensive discussions in astrophysics as well as particle and nuclear physics. In particular, the discovery of the neutrino oscillation is one of the most outstanding achievements in the modern physics \cite{Kajita:2016cak, McDonald:2016ixn}.

However, still the unexplained remains; neutrino anomalies have been reported, which are not partly in accord with the 3 neutrino model \cite{Aguilar:2001ty, AguilarArevalo:2010wv, Bhattacharya:2011ah, Giunti:2010zu}. Although some ambiguities related to nuclear physics in reactors are being intensively investigated in theoretical and experimental sides \cite{Dwyer:2014eka, Ashenfelter:2018iov, Petkovic:2019wyw}, the inactive fourth flavor neutrino called ``sterile neutrino ($\nu_s$)" is still intriguing as one of possibilities to explain the anomalies and some other peculiar phenomena related to neutrinos \cite{Boyarsky:2018tvu}.

Astrophysics studies on big bang nucleosynthesis \cite{Archidiacono:2013xxa, Hamann:2010bk, Mangano:2011, Jang:2016rpi} and X-ray observations from galaxy clusters \cite{Chan:2013uu, Boyarsky:2006zi, Boyarsky:2014jta, Bulbul:2014sua} constrain the properties of $\nu_s$ and propose it as a dark matter candidate. In addition, SN studies found that the $\nu_s$ can enhance the SN explosion energy using the constraints on the mixing parameter from the X-ray astronomy \cite{Warren:2014qza}. Furthermore, it is investigated that active-sterile ($\nu_a$-$\nu_s$) flavor conversion in core-collapse SN leads to enhancing the r-process elements so as to coincide with the observation of the metal-poor star HD 122563 \cite{Wu:2013gxa}.  

Although the astrophysical phenomena leave a room for a possible existence of $\nu_s$, IceCube experiments exclude a parameter region for the mixing between $\nu_s$ and $\nu_\mu$ indicated by shortbase line (SBL) experiments, by showing no evidence of the anomalous $\nu_\mu$-disappearance \cite{TheIceCube:2016oqi}. In the analysis, IceCube collaborations assumed $\theta_{14} = \theta_{34} =0$ to take into account the $\nu_\mu$ disappearance, whereas the condition of $\sin^2 2 \theta_{\mu e} \neq 0$ in SBL anomalies requires the $\theta_{14} \neq 0$. By improving the inconsistency, a recent work finds a new global fitted mixing parameters of the 3+1 neutrino model as $\Delta m_{41}^2 \approx 1.75\,\eV^2$, $\theta_{14} = 9.44^{\circ}$,  $\theta_{24} = 6.93^{\circ}$ and $\theta_{34} = 0$ \cite{Collin:2016aqd}. In this letter, we adopt those results to study the effects of $\nu_a$-$\nu_s$ neutrino oscillation on the SN neutrino process ($\nu$-process). 

Despite the small cross section between neutrinos and nuclei, the $\nu$-process turns out to be important to explore origin of some rare nuclei. For instance, the neutrino-induced nucleosynthesis in SN significantly affects abundances of light-to-heavy nuclei such as $^{7}$Li, $^{11}$B, $^{92}$Nb, $^{98}$Tc, $^{138}$La and $^{180}$Ta \cite{Woosley:1989bd, Heger:2003mm,Hayakawa:2018ekx}. In order to include the $\nu_s$ in the $\nu$-process in SN, we apply the 3+1 neutrino model and evaluate their viability. As a source of $\nu_s$s, we consider two models for production mechanism\,\cite{Kolb:1996pa}. First, we assume the non-interacting $\nu_s$s produced only by mixing with $\nu_a$s. Second, $\nu_s$s can be directly created by electron neutrino collisions with matter. 

One of the most important physical inputs in the $\nu$-process is how to describe the neutrino oscillation behavior in SN environments. Generally, the time evolution of the 3+1 neutrino model is expressed as the following  Schr\"odinger-like equation 
\begin{eqnarray}
i{\rm{\frac{d}{dt}}} \begin{pmatrix} \nu_e \\ \nu_\mu \\ \nu_\tau \\ \nu_s \end{pmatrix} = \hat{H} \begin{pmatrix} \nu_e \\ \nu_\mu \\ \nu_\tau \\ \nu_s \end{pmatrix},
\label{eq1}
\end{eqnarray}
where we use units of $\hbar = c \equiv 1$. For the non-interacting $\nu_s$, we compose the total Hamiltonian as vacuum ($\hat{H}_{\rm vacuum}$) and matter ($\hat{H}_{\rm matter}$) terms,
\begin{eqnarray}
&&\hat{H}_{\rm vacuum} = U {\rm diag}(0, \frac{\Delta m_{21}^2}{2E_\nu}, \frac{\Delta m_{31}^2}{2E_\nu}, \frac{\Delta m_{41}^2}{2E_\nu} ) U^{\dagger},  \\[12pt] \nonumber 
&& \hat{H}_{\rm matter} = {\rm diag} (V_{\rm CC} +V_{\rm NC}, V_{\rm NC}, V_{\rm NC}, 0),
\label{eq2}
\end{eqnarray} 
where $U$ and $E_\nu$, respectively, denote the $4\times 4$ unitary mixing matrix \cite{Collin:2016aqd} and neutrino energy. The mass squared difference is defined as $\Delta m_{ij}^2 \equiv m_i^2 - m_j^2$. Mixing parameters for three active flavors are adopted from the Particle Data Group \cite{Agashe:2014kda}. Mass eigenstates are composed of the highest fourth mass state dominated by the $\nu_s$ \cite{Collin:2016aqd} and others for three $\nu_a$s whose mass hierarchy (MH) has two possible cases known as normal and inverted hierarchies (NH and IH). In the non-interacting $\nu_s$ model, the fourth diagonal term in $\hat{H}_{\rm matter}$ would be zero while other terms for active flavors are given as charged and neutral current potentials ($V_{\rm CC}$ and $V_{\rm NC}$). Note that the fourth term can also be neglected for the second model since the interaction strength of $\nu_s$ is constrained to be weak enough \cite{Kolb:1996pa, Barbieri:1988av}.  

With the 3+1 neutrino model, we investigate the SN $\nu$-process up to the He layer with Lagrange mass coordinate ranging from 1.6 $M_{\odot}$ to 6 $M_{\odot}$, which is based on SN1987A model with the metallicity in Large Magellanic Cloud (LMC) $i.e.$ $Z = 1/4 ~ Z_\odot$. We use an explosion energy of $10^{51}\,{\rm erg}$ and a hydrodynamical model in Ref. \cite{Kusakabe:2019znq} based on the public code, blcode \cite{Ott:blcode}. The reaction network and thermal nuclear reactions rates are, respectively, taken from Ref.\,\cite{Kusakabe:2019znq} and JINA data base \cite{Cyburt:2010}. For cross sections of neutrino-induced reactions, those for $^4$He and $^{12}$C are adopted from Ref.\,\cite{Yoshida:2008zb}; those for intermediate nuclei of $^{12}$C to $^{80}$Kr from \cite{HW92}; those for heavy elements related to production of $^{92}$Nb, $^{98}$Tc, $^{138}$La and $^{180}$Ta from \cite{Cheoun:2010, Cheoun:2011hj}. 

\begin{figure}[t]
\includegraphics[width=8.6cm]{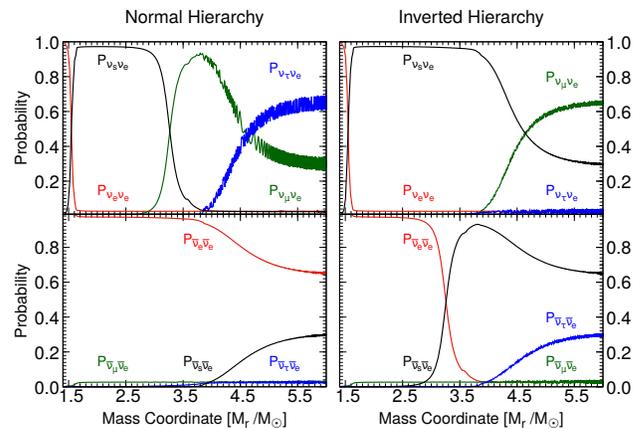}
\caption{Survival probability of $\nu_e$ with $E_\nu=$15\,MeV as a function of mass coordinate.   $P_{\nu_\alpha\nu_e}$ denotes the flavor change probability from $\alpha$ to $e$ flavor. Left and right panels show results for the NH and IH, respectively. Upper and lower panels, respectively, describe the probability $P_{\nu_\alpha\nu_e}$ for neutrinos and anti-neutrinos.}
\label{fig1}
\end{figure}

Figure\, \ref{fig1} shows the survival probability of $\nu_e$ in SN explosion with the 3+1 neutrino model. For the NH case (left-upper panel), we find three kinds of resonances resulting from the Mikheyev--Smirnov--Wolfenstein effect \cite{Wolfenstein:1977ue, Mikheyev1986}. For the resonance condition, the larger the mass squared difference is, the higher matter density is required. Therefore, the conversion between $\nu_e$ and $\nu_s$ occurs in the first resonance region at $M_r \sim 1.6 M_{\odot}$ since $\Delta m^2_{41}$ is the largest and the \nth{4} mass state is dominantly occupied by $\nu_s$. Note that this resonance region is moved to inner side when sterile neutrino is heavier, but the neutrino process is not significantly affected by this change. The second resonance follows in the regions of $3 M_{\odot} < M_r < 3.5 M_{\odot}$, where $\nu_\mu$ converts to $\nu_e$ by the similar condition. In the case of the IH, the conversion between the $\nu_s$ and $\nu_e$ also occurs in the similar region. The difference is in the outer region where the probability $P_{\nu_s \nu_e}$ survives instead of $P_{\nu_\tau \nu_e}$ value in NH. 

The oscillation probability largely affects the neutrino-induced reaction rates. For the 3+1 neutrino model, the neutrino reaction rate can be written as   
\begin{eqnarray}
\lambda (t,r;T_{\nu_\beta}) =  
\frac{1}{4\pi r^2}
\sum_{\substack{ \alpha,\beta = \\   e,\mu, \tau, s}}
\frac{{\cal L}_{\nu_\beta}(t)}{\left\langle E_{\nu_\beta} \right\rangle} 
\left\langle \sigma_{\nu_\alpha} P_{\nu_{\beta} \nu_{\alpha}}(r;T_{\nu_\beta}) \right\rangle ~~~~
\label{eq3}
\end{eqnarray}
where $r$ is the radius from the center of proto-neutron star. $\left\langle E_{\nu_\beta} \right\rangle$ and $\left\langle \sigma_{\nu_\alpha} P_{\nu_{\beta} \nu_{\alpha}}\right\rangle$, respectively, denote the thermally averaged energy and the averaged cross section multiplied by the flavor change probability with Fermi-Dirac distribution for neutrino temperature $T_{\nu_{\beta}}$ used in Ref.\ \cite{Hayakawa:2018ekx}. $\mathcal{L}_{\nu_\beta}$ stands for the luminosity of the $\nu_\beta$ determined by initial conditions.

In general, the $\mathcal{L}_{\nu_s}$ depends on their production rate, which is determined by the oscillation probability ($P_{\nu_a \nu_s}$) for the non-interacting $\nu_s$. Near the SN core, the high baryon density $(\rho_B \sim 10^{14} {\rm g/cm^3})$ strongly suppresses the probability $P_{\nu_a \nu_s}$, and $\mathcal{L}_{\nu_s}$ is feeble in comparison with $\mathcal{L}_{\nu_a}$. Hence, in the first model, we set the initial $\mathcal{L}_{\nu_s}$ to zero (termed as null ${\cal L}_{\nu_s}$ model), while $\mathcal{L}_{\nu_a}$ is the same as the 3 neutrino model \cite{Yoshida:2005uy}. Recent calculation of the $\mathcal{L}_{\nu_a}$ by $\nu$-transport simulation \cite{OConnor:2018sti} may affect the reaction rate, but the neutrino oscillation behavior is changed rarely.

\begin{figure}[t]
\centering
\includegraphics[width=8.6cm]{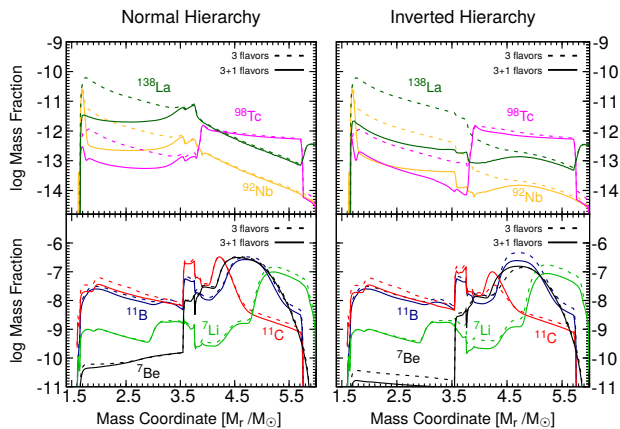}
\caption{Mass fractions of $^7$Li, $^7$Be, $^{11}$B, $^{11}$C, $^{92}$Nb, $^{98}$Tc and $^{138}$La as a function of the mass coordinate after 50 seconds from SN explosion. Left and right panels, respectively, correspond to the NH and IH cases. Dashed and solid lines denote the results in the 3 and the 3+1 neutrino model with the null $\mathcal{L}_{\nu_s}$ model, respectively.}
\label{fig2}
\end{figure}

Figure\,\ref{fig2} shows nuclear abundances of light-to-heavy elements produced by SN $\nu$-process with the 3 and the 3+1 neutrino model by using the null ${\cal L}_{\nu_s}$ model. The heavy nuclei such as $^{92}$Nb, $^{98}$Tc and $^{138}$La are mainly produced at $1.5 M_{\odot} < M_r < 3.5 M_{\odot}$. In this region, regardless of MH, the conversion of $\nu_e \rightleftharpoons \nu_s$ suppresses the flux of $\nu_e$ because of the null ${\cal L}_{\nu_s}$ condition. The suppression of $\nu_e$ reduces the rates of the main reactions $^{92}$Zr($\nu_e, e^-)^{92}$Nb, $^{98}$Mo($\nu_e, e^-)^{98}$Tc and $^{138}$Ba($\nu_e, e^-)^{138}$La. Consequently, mass fractions of the heavy nuclei are decreased in this model. In the region of $M_r > 3.5 M_{\odot}$, on the other hand, abundances of heavy elements in the 3+1 neutrino model of IH case are smaller than those by the NH and 3 neutrino model. This result comes from the incomplete recovery of $\nu_e$ in the outer region (Fig.\,\ref{fig1}). $\nu_\mu$ and $\nu_\tau$ cannot sufficiently replenish the $\nu_e$ flux in the outer region in the IH case, although the flavor change to $\nu_e$ occurs at $M_r \gtrsim 4M_{\odot} $. Therefore, unlike the 3 neutrino model, total abundances of heavy elements in the 3+1 neutrino model depend on the neutrino MH.

In the outer region, the different oscillation patterns between NH and IH significantly affect light elements abundances (lower panels in Fig.\,\ref{fig2}). For the NH case, the suppression of $\bar{\nu}_e$ by $\bar{\nu}_s$ reduces the $^7$Li abundance because of a decreased rate of $^4$He($\bar{\nu}_e, e^+ n$)$^3$H($\alpha,\gamma$)$^7$Li. For the IH case, on the other hand, the reaction $^4$He$(\nu_e,e^-p)^3$He as well as $^4$He($\bar{\nu}_e, e^+ n$)$^3$H is hindered by a suppressed flux of $\nu_e$. Thus, the reduced fluxes of $\nu_e$ and $\bar{\nu}_e$ result in significantly fewer $^3$H and $^3$He (compared with those in the 3 neutrino model \cite{Kusakabe:2019znq}). This leads to reduced final abundances of $^{7}$Li and $^{11}$B which are produced via radiative $\alpha$ capture reactions.

Table \ref{table1} shows the abundance ratio of $^{7}$Li and $^{11}$B by the null ${\cal L}_{\nu_s}$ model with the observational data from the bayesian analysis of SiC X grains. We assume that elements in SiC X grains are uniformly mixed before it condensed in the SN ejecta long before the solar system formation. With the assumption, we integrate the mass fractions of $^{7}$Li and $^{11}$B after the decay of unstable nuclei over the whole mass region. The analysis of SiC X grains constrains the ratio of $^{7}$Li and $^{11}$B produced by the neutrino process as $^{7}$Li/$^{11}$B$=-0.31 \pm 0.42$ \cite{Mathews:2011jq}. Our results by the null ${\cal L}_{\nu_s}$ model demonstrate that the 3 neutrino model is only allowed in the IH case within the $3\sigma$ limit and the 3+1 neutrino model is excluded. 

As the second model, we consider that $\nu_s$s can be directly produced in $\nu_e$ collision with matter. In this case, the cross section for the $\nu_s$ production can be written as $\sigma(\nu_e X \rightarrow \nu_s X) = A \sigma_{eX} (\nu_e X \rightarrow \nu_e X)$, where $X$ is an electron, neutron, or proton, and $A$ is a parameter for the interaction strength. The cross section leads to the luminosity of $\nu_s$ as \cite{Kolb:1996pa}
\begin{eqnarray}
\mathcal{L}_{\nu_s} \simeq n_e n_{\nu_e} A \sigma_{ee} V \left\langle E \right\rangle, 
\label{eq4}
\end{eqnarray}
where $n_i$ is the number density of particle $i$, $V$ is the volume of the core, and $\left\langle E \right\rangle$ is an average energy of  the neutrino. According to Ref.\,\cite{Kolb:1996pa}, total neutrino luminosity of SN1987A puts the constraint $A \le 10^{-10}$. When we choose the upper limiet $A = 10^{-10}$ and the core temperature of $T_C = 50 \, {\rm MeV}$, $\mathcal{L}_{\nu_s}$ becomes about $10^{51} \rm erg/s$, which can be comparable with $\mathcal{L}_{\nu_a}$. If the $\mathcal{L}_{\nu_s}$ is not negligible in this way, energetic $\nu_a$s converted from the $\nu_s$s increase relevant neutrino reaction rates because earlier decoupling of feebly interacting $\nu_s$s may result in a higher temperature of decoupled $\nu_s$s.
\begin{table}[t]
	\centering
    \begin{tabular}{c|c|c|lc}
        \hline
		\hline
		~ $\nu$ model & ~NH & ~IH & ~Observation \cite{Mathews:2011jq} & ~  \\
		\hline
		~ 3 flavors 										& ~$1.22$~	& ~$0.80$		& ~$< 0.53 \, (2 \sigma\ 95\%\ {\rm C.L.} )$	    \\
            \cline{1-3}		
		~ 3+1 flavors    	& ~$1.27$	& ~$1.04$	& ~$< 0.95 \, (3 \sigma\ 99.7\%\ {\rm C.L.} )$		  \\
		\hline
		\hline
	\end{tabular}
		\caption{Abundance ratio of $^{7}$Li and $^{11}$B produced within the whole mass coordinate region in the null $\mathcal{L}_{\nu_s}$ model and observations from analysis of SiC X grains.}
		\label{table1}
\end{table}
As a simple test, we assume that the $\nu_s$ has an equivalent luminosity with those of $\nu_a$s, {\it i.e.}, $\mathcal{L}_{\nu_s} =  \mathcal{L_\nu}/8$ (termed as an equivalent ${\cal L}_{\nu_s}$ model), where $\mathcal{L}_{\nu}$ is the total neutrino luminosity given as $10^{53}\,{\rm erg/s}$. Also, we set the temperature of the $\nu_s$ ($T_{\nu_s}$) as a free parameter. The exact $\mathcal{L}_{\nu_s}$ and temperature should be derived by treating the decoupling process of $\nu_s$s with the Boltzmann equation. It is, however, beyond the scope of the present paper.
 
In this equivalent luminosity model, the higher temperature the $\nu_s$ has, the more energetic $\nu_a$s are produced by the neutrino oscillation. The energetic neutrinos enhance the neutrino reaction rates in Eq.\,(3). As a result, abundances of the heavy nuclei, {\it i.e.}, $^{92}$Nb, $^{98}$Tc and $^{138}$La, are increased, contrary to the trend in Fig.\,\ref{fig2} for the null $\mathcal{L}_{\nu_s}$ model. 

Table \ref{table2} shows the ${^7}$Li/$^{11}$B ratio with the equivalent $\mathcal{L}_{\nu_s}$ model. In the NH case, the conversion of $\nu_s$s ($\bar{\nu}_s$s) with high temperature enhances rates of $^4$He($\bar{\nu}_e, e^+ n$)$^3$H and $^4$He(${\nu}, \nu'p$)$^3$H, which leads to an increased $^{11}$B abundance. On the other hand, the final $^{7}$Li yield is predominantly contributed by the $^7$Be produced via $^4$He($\nu_e, e^- p$)$^3$He($\alpha, \gamma$)$^7$Be and $^4$He($\nu, \nu'n$)$^3$He($\alpha,\gamma$)$^7$Be. While the $\nu_e$ reaction rates are not significantly increased in the NH case, $\nu_e$s produced by $\nu_s$s in the IH case enhance the $\nu_e$ reaction rates. Besides, in the IH case, more $\bar{\nu}_e$s converted from $\bar{\nu}_s$s also contribute to the increase of the $^{7}$Li abundance. Therefore, the $^7$Li/$^{11}$B ratio is smaller in the NH case and larger in the IH case for higher $\nu_s$ temperatures.

The analysis data from SiC X allows both MHs in the 3+1 neutrino model within the $3\sigma$ limit for $T_{\nu_s}=7\,{\rm MeV}$, whereas the only NH case is allowed for $T_{\nu_s} \ge 8 \,{\rm MeV}$. When we consider that the typical core temperature of proto-neutron stars is about 50\,MeV, $\nu_s$s can carry much higher temperature than the highest adopted value in our calculation. Therefore, we can expect that the SiC X analysis may exclude the IH case of this model. 

\begin{table}[t]
 	\centering
	\begin{tabular}{c|cccc}
        \toprule
        \multicolumn{1}{c|}{Neutrino}  & \multicolumn{4}{c}{Temperature of ${\nu_s}$} \\
        Mass Hierarchy &\ 7\,MeV\ &\ 8\,MeV\ &\ 9\,MeV\ &\ 10\,MeV\  \\
		\hline
		
		~NH	    & ~$0.88$~	& ~$0.82$	& ~$0.78$~ & ~$0.78$~  \\
		~IH 	& ~$0.92$~	& ~$1.02$	& ~$1.15$~ & ~$1.15$~  \\
		\hline
        \hline
	\end{tabular}
		\caption{Total abundance ratio of $^{7}$Li and $^{11}$B in the 3+1 neutrino model with the equivalent luminosity of all flavors. }
		\label{table2}
\end{table}
In summary, the present work investigates the viability of the 3+1 neutrino model in the SN $\nu$-process adopting the global fitting data of IceCube and SBL experiments. For the 3+1 neutrino model, two luminosity models are investigated. First, for the non-interacting $\nu_s$s, it is unlikely to produce the $\nu_s$s in the inner region due to the small mixing, so we assume the null $\mathcal{L}_{\nu_s}$. The flavor change from $\nu_e$ to $\nu_s$ reduces neutrino reaction rates in the SN $\nu$-process, and the SiC X analysis excludes the 3+1 neutrino model at the $3 \sigma$ level. Second, we take the equivalent luminosity for all flavors. In that case, the neutrino oscillation enhances neutrino reaction rates due to the high temperature of decoupled $\nu_s$s. Comparing our result with the SiC X analysis, it is shown that the 3+1 neutrino model is only allowed at $T_{\nu_s}=7\,{\rm MeV}$ in the IH case within the $3 \sigma$ limit, while the NH case is fully allowed. Even though we regard $T_{\nu_s}$ as the free parameter in this calculation, it is important to estimate the $T_{\nu_s}$ and $\mathcal{L}_{\nu_s}$ in the SN explosion based upon an accurate kinetic theory for the neutrino transport. We expect that future study on the decoupling process of $\nu_s$s would suggest a clue for the existence of $\nu_s$s as well as the neutrino MH. 

This work of H.K., D.J. and M.-K.C. was supported by Korea National Research Foundation of Korea (Grants No.\ 2013M7A1A10757b4 and No.\ NRF-2017R1E1A1A01074023). The work of M.K. was supported by NSFC Research Fund for International Young Scientists (11850410441).

\end{document}